\def\sorb{b}
\if s\sorb
  \documentstyle{article}
  \newlength{\absize}
  \renewcommand{\baselinestretch}{1.0}
  \renewcommand{\arraystretch}{0.5}
  
\begin{document}
  \date{}
  \pagestyle{empty}
  \thispagestyle{empty}
  \renewcommand{\thefootnote}{\fnsymbol{footnote}}
  \newcommand{\starttext}{\newpage\normalsize
    \pagestyle{plain}
    \setlength{\baselineskip}{4ex}\par
    \twocolumn\setcounter{footnote}{0}
    \renewcommand{\thefootnote}{\arabic{footnote}}}
\else
  \documentstyle[12pt,a4wide,epsf]{article}
  \newlength{\absize}
  \setlength{\absize}{\textwidth}
  \renewcommand{\baselinestretch}{1.0}
  \renewcommand{\arraystretch}{0.5}
  \begin{document}
  \thispagestyle{empty}
  \pagestyle{empty}
  \renewcommand{\thefootnote}{\fnsymbol{footnote}}
  \newcommand{\starttext}{\newpage\normalsize
    \pagestyle{plain}
    \setlength{\baselineskip}{4ex}\par
    \setcounter{footnote}{0}
    \renewcommand{\thefootnote}{\arabic{footnote}}}
\fi
\newcommand{\preprint}[1]{%
  \begin{flushright}
    \setlength{\baselineskip}{3ex} #1
  \end{flushright}}
\renewcommand{\title}[1]{%
  \begin{center}
    \LARGE #1
  \end{center}\par}
\renewcommand{\author}[1]{%
  \vspace{2ex}
  {\Large
   \begin{center}
     \setlength{\baselineskip}{3ex} #1 \par
   \end{center}}}
\renewcommand{\thanks}[1]{\footnote{#1}}
\renewcommand{\abstract}[1]{%
  \vspace{2ex}
  \normalsize
  \begin{center}
    \centerline{\bf Abstract}\par
    \vspace{2ex}
    \parbox{\absize}{#1\setlength{\baselineskip}{2.5ex}\par}
  \end{center}}

\setlength{\parindent}{3em}
\setlength{\footnotesep}{.6\baselineskip}
\newcommand{\myfoot}[1]{%
  \footnote{\setlength{\baselineskip}{.75\baselineskip}#1}}
\renewcommand{\thepage}{\arabic{page}}
\setcounter{bottomnumber}{2}
\setcounter{topnumber}{3}
\setcounter{totalnumber}{4}
\newcommand{\figsize}{}
\renewcommand{\bottomfraction}{1}
\renewcommand{\topfraction}{1}
\renewcommand{\textfraction}{0}
\newcommand{\beq}{\begin{equation}}
\newcommand{\eeq}{\end{equation}}
\newcommand{\beqa}{\begin{eqnarray}}
\newcommand{\eeqa}{\end{eqnarray}}
\newcommand{\nn}{\nonumber}

\newcommand{\dd}{\mbox{{\rm d}}}
\newcommand{\mH}{m_{\rm H}}
\newcommand{\dLips}{\mbox{{\rm dLips}}}

%
\newcommand{\GeV}{\mbox{{\rm GeV}}}
\newcommand{\gs}{g_{\rm s}}
\renewcommand{\Re}{{\rm Re}}
\renewcommand{\Im}{{\rm Im}}
\def\gsim{\mathrel{\rlap{\raise 2.5pt \hbox{$>$}}\lower 2.5pt
\hbox{$\sim$}}}
\def\lsim{\mathrel{\rlap{\raise 2.5pt \hbox{$<$}}\lower 2.5pt
\hbox{$\sim$}}}

\newcommand{\postscript}[2]
{\setlength{\epsfxsize}{#2\hsize}
\centerline{\epsfbox{#1}}}
\def\slash#1{#1 \hskip -0.5em /}
%
%

\preprint{University of Bergen, Department of Physics \\
Scientific/Technical Report No.\ 1994-14 \\ ISSN~0803-2696\\
November 1994}

\vfill
\title{PROBING $CP$ IN THE HIGGS SECTOR \thanks{
                 To appear in: Proceedings of the conference
                 {\it IX Workshop on High
                 Energy Physics and Quantum Field Theory}, Zvenigorod,
                 Russia, September 16-22, 1994}}

\vfill
\author{Arild Skjold \thanks{Electronic mail address:
                {\tt skjold@vsfys1.fi.uib.no}}\\\hfil\\
        Department of Physics \\
        University of Bergen \\ All\'egt.~55, N-5007 Bergen, Norway }
\date{}

\vfill
\abstract{We discuss how one may determine whether some Higgs particle is an
eigenstate of $CP$ or not, and if so, how we may probe
whether it is $CP$-odd or $CP$-even. The idea is applied to Higgs decay where
correlations
among momenta of the decay products may yield the desired information.
These are correlations of decay planes defined by the momenta of pairs of
particles as well as correlations between energy differences.
In the case that the Higgs particle is an eigenstate of $CP$, our study
includes
finite-width effects.
The correlations between energy differences turn
out to be a much better probe for investigating $CP$ properties than the
previously suggested angular correlations, especially for massive Higgs bosons.
}

\vfill

\starttext

\renewcommand{\theequation}{\thesection.\arabic{equation}}
\section{Introduction}
\label{sec:intro}
\setcounter{equation}{0}

When some candidate for the Higgs particle is discovered, it becomes
imperative to establish its properties, other than the mass.
While the standard model Higgs boson is even under $CP$,
alternative and more general models contain Higgs bosons
for which this is not the case. One may even have
$CP$ violation present in the Higgs sector.
We shall here discuss how decay distributions may disentangle a scalar Higgs
candidate from a pseudoscalar one \cite{osskj1}, and next, by allowing for
$CP$ violation in the Higgs sector, discuss some possible signals of such
effects \cite{osskj2}.
This situation is similar to the classical one of determining
the parity of
the $\pi^0$ from the angular correlation of the planes spanned by the
momenta of the two Dalitz pairs \cite{Yang50},
$\pi^0\rightarrow\gamma\gamma\rightarrow(e^+e^-)(e^+e^-)$.

In non-standard or extended models of the electroweak interactions,
there exist ``Higgs-like'' particles having negative $CP$.
An example of such a theory is
the minimal supersymmetric model ($MSSM$) \cite{HHG}, where
there is a neutral $CP$-odd Higgs boson, often denoted $A^{0}$ and sometimes
referred to as a pseudoscalar.

The origin of mass and the origin of $CP$ violation are widely considered
to be the most fundamental issues in contemporary particle physics.
Perhaps for this very reason, there has been much speculation
on a possible connection.  This is also in part caused by
the fact that the amount of $CP$ violation accommodated in the CKM matrix
does not appear sufficient to explain the observed baryon
to photon ratio \cite{Cohen}.

We shall here consider the possibility that $CP$ violation is present
in the Higgs sector, and discuss some possible signals of such effects
in Higgs decay.
While the standard model induces $CP$ violation in the Higgs sector at
the one-loop level provided
the Yukawa couplings contain both scalar and pseudoscalar components
\cite{Wein}, we actually have in mind
an extended model, such as e.g., the two-Higgs-doublet model
\cite{TDLee}.

Below we postulate an effective Lagrangian which contains $CP$ violation in
the Higgs sector.
In cases considered in the literature, $CP$ violation usually
appears as a one-loop effect. This is due to the fact that the $CP$-odd
coupling introduced below
is a higher-dimensional operator and in renormalizable models these are
induced only at loop level. Consequently we expect the
effects to be small and any observation of $CP$ violation to be
equally difficult.

$CP$ non-conservation has manifested itself so far only in the neutral kaon
system. In the context of the Standard Model this $CP$ violation
originates from the Yukawa sector via the CKM matrix \cite{KoMa}. Although
there may be several sources of $CP$ violation, including the one above, we
will here consider a simple model where the $CP$ violation is restricted
to the Higgs sector and in particular to the coupling
between some Higgs boson and the vector bosons.
Specifically, by assuming that the coupling between the Higgs boson $H$
and the vector bosons $V=W, Z$ has both scalar
and pseudoscalar components,
the most general
coupling for the $HVV$-vertex relevant for decays to massless
fermions may be written as
\cite{Nel,Cha}
\beq
i\ \left(2\cdot 2^{1/4}\right) \sqrt{G_{\rm F}}
\left[ m_V^2 \ g^{\mu \nu}
+ \xi\left(k_1^2,k_2^2\right)
\ \left(k_{1} \cdot k_{2} \ g^{\mu \nu} - k_{1}^{\nu} k_{2}^{\mu} \right)
+ \eta\left(k_1^2,k_2^2\right)
\ \epsilon^{\mu \nu \rho \sigma} k_{1 \rho} k_{2 \sigma} \right],
\label{EQU:int2}
\eeq
with $k_{j}$ the momentum of vector boson $j$, $j=1,2$.
The first term is the familiar $CP$-even $V^\mu V_\mu H$ tree-level $SM$
coupling. The second term stems from the dimension-5 $CP$-even operator
$V^{\mu\nu} V_{\mu\nu} H$ with
$V_{\mu \nu}=\partial_\mu V_\nu -\partial_\nu V_\mu$. The last term is $CP$ odd
and originates from the dimension--5 operator $\epsilon^{\mu \nu \rho \sigma}\
V_{\mu \nu} V_{\rho \sigma} \ H$.
Simultaneous presence of $CP$-even and $CP$-odd terms leads to $CP$ violation,
whereas presence of only the $CP$ odd term describes a pseudoscalar.
The higher-dimensional operators are radiatively induced and we may therefore
safely neglect the contribution from the second term. The strength parameter
$\eta$ may in general be complex, with the imaginary part arising from e.g.
final state interactions.

Related studies have been reported by \cite{Nel} in the context
of correlations between decay planes involving scalar and (technicolor)
pseudoscalar Higgs bosons.
It should be noted that the present discussion is more general,
since finite-width effects of the vector bosons are taken into account.
This enables us to investigate the angular correlations for Higgs masses below
the threshold for decay into real vector bosons.
In the context of $CP$ violation, related studies have been reported by
\cite{Cha}. Recently, Arens et.al.~\cite{Arens} have investigated the energy
spectrum in Higgs decay to four fermions. In doing so, one does not have to
reconstruct the decay planes, and hence one is able to study final states
involving neutrinos and identical fermion pairs.
In both cases we discuss correlations between energy differences. It
turns out that under suitable experimental conditions these correlations
provide an even better signal for investigating $CP$ properties.

In the next section we will discuss the possibilities of distinguishing
$CP$-even from $CP$-odd eigenstates. In section~3, we will study possible
signals of $CP$ violation in Higgs decay.

\section{Signals of $CP$ eigenstates in Higgs decay}
\setcounter{equation}{0}
\label{sec:eigen}
We consider here the decay of a standard-model Higgs ($H$) or
a `pseudoscalar' Higgs particle ($A$), via two vector
bosons ($W^+W^-$ or $ZZ$), to two
non-identical fermion-anti-fermion pairs,
\begin{eqnarray}
\label{EQU:Hdecay}
(H,A)\rightarrow V_1V_2\rightarrow (f_1\bar f_2)(f_3\bar
f_4).
\end{eqnarray}
The two vector bosons need not be on mass shell.

If we let the momenta ($q_1$, $q_2$, $q_3$, and $q_4$)
of the two fermion-antifermion pairs
(in the Higgs rest frame) define two planes,
and denote by $\phi$ the angle between those two planes,
then we shall discuss the angular distribution of the decay rate
$\Gamma$,
\beq
\frac{1}{\Gamma}\:
\frac{{\mbox{{\rm d}}\Gamma}}{\mbox{{\rm d}}\phi}
\label{EQU:intro1}
\eeq
and a related quantity, to be defined below, both
in the case of $CP$-even and $CP$-odd Higgs bosons.

In order to parameterize the vector and axial couplings, we define
the couplings involving vector bosons $V_1$ and $V_2$
in terms of angles $\chi_1$ and $\chi_2$ as
\beq
g^{(i)}_{V}  \equiv  g_i \cos \chi_i, \qquad
g^{(i)}_{A} \equiv g_i \sin \chi_i, \qquad i=1,2.
\label{EQU:Dj9}
\eeq
The only reference to these angles is through $\sin2\chi$.
Relevant values are given in table~1 of ref.~\cite{osskj1}.
The couplings of $H$ and $A$ to the vector bosons are given by retaining only
the first and last term in (\ref{EQU:int2}), respectively.

We first turn to a discussion of angular correlations.
The distributions of
eq.~(\ref{EQU:intro1}) take the form
\beqa
\frac{2 \pi}{\Gamma_{H}}\:\frac{\dd\Gamma_{H}}{\dd\phi}
& = &
1 + \alpha\left(m\right)\sin(2 \chi_{1})\sin(2 \chi_{2}) \cos \phi
+ \beta\left(m\right)\cos 2 \phi ,
\label{EQU:Dl5} \\
\frac{2 \pi}{\Gamma_{A}}\:\frac{\dd\Gamma_{A}}{\dd\phi}
& = &
1 - \frac{1}{4} \cos 2 \phi .
\label{EQU:Dl6}
\eeqa
The functions $\alpha$ and $\beta$ depend on
the ratios of the masses of the vector bosons to the Higgs mass.
They are given in \cite{osskj1}.
In the narrow-width approximation these decay correlations
are identical to the ones reported in \cite{Nel} and our
analysis fully justifies this approximation.

However, our analysis is valid also below the threshold
for producing real vector bosons, $m < 2 \, m_{V}$.
A representative set of angular distributions are given
in Fig.\ref{plfun2} (left) for the cases
$H \rightarrow (l^{+} \nu_{l}) (b \overline{c})$
and $H \rightarrow (l^{+} l^{-}) (b \overline{b})$ for
$m=70, 150, 300$~GeV and $m=70, 300$~GeV, respectively.
(Of course, jet identification is required for this kind of analysis.)
There is seen to be a clear difference between the $CP$-even
and the $CP$-odd cases.
\begin{figure}[t]
\begin{center}
\setlength{\unitlength}{1cm}
\begin{picture}(16,7.0)
\put(-0.5,-2.5){\mbox{\epsfysize=11cm\epsffile{diffwa.eps}}
             \hspace*{-10mm}
             \mbox{\epsfysize=11cm\epsffile{diffk.eps}}}
\end{picture}
\vspace*{15mm}
\end{center}
\caption{\label{plfun2}{\it Left} : Angular distributions of the
planes
defined by two fermion pairs for $CP$-even Higgs particles decaying via two
$W$'s and two $Z$'s, compared with the corresponding distribution
for the $CP$-odd case (denoted $A$).
{\it Right }: The energy-weighted case of eq.~(\ref{EQU:Dl7}).}
\end{figure}

Let us now turn to a discussion of energy differences. In order to make the
difference between the $SM$ and $CP$ odd distributions more significant, we
multiply the differential cross sections with the energy differences
$(\omega_{1}-\omega_{2})(\omega_{3}-\omega_{4})$ before
integrating over energies. The corresponding energy-weighted distribution
takes the form
\beq
\frac{2 \pi}{\tilde{\Gamma}}
\:\frac{\dd\tilde{\Gamma}}{\dd\phi}
 =
1 + \frac{\kappa\left(m\right)}{\sin(2 \chi_{1})\sin(2 \chi_{2})}
\cos \phi,
\label{EQU:Dl7}
\eeq
in the $CP$-even case, whereas there is no correlation in the $CP$-odd case.
The
function $\kappa$ is given in \cite{osskj1}.
A representative set of distributions
(\ref{EQU:Dl7}) are given in Fig.\ref{plfun2} (right)
in the cases $H \rightarrow (l^{+} \nu_{l})
(b \overline{c})$ and $H \rightarrow (l^{+} l^{-})
(b \overline{b})$ for $m=70, 300, 500$~GeV and $m=70, 300$~GeV,
respectively. In this latter case, we compare an uncorrelated
distribution with a strongly correlated one.

\section{Signals of $CP$ violation in Higgs decay}
\setcounter{equation}{0}
\label{sec:viol}

The present discussion will be along similar lines as in the preceeding
section, now allowing for both the $SM$ and the $CP$ odd term in
(\ref{EQU:int2}). Although $\eta$ may be complex in general, we
shall here assume it to be real. The reasons are threefold: 1) The results turn
out to be valid for any value of $\eta$. 2) The predictive power increases by
introducing one, and not two, parameters. 3) If we had kept $\eta$
complex our results would still have been valid to order
$\left(\mbox{\rm Im\ }\eta\right)^{2}$ in the case of purely angular
distributions and in the energy-weighted case the imaginary term would have
been suppressed by $\sin2\chi$-factors.

Due to our ignorance concerning
$\eta$, we have to use the narrow-width approximation.
This is of course only meaningful above threshold for producing real
vector bosons.
The distribution of eq.~(\ref{EQU:intro1}) then takes the compact form
\beq
\frac{2 \pi}{\Gamma}\:\frac{\dd\Gamma}{\dd\phi}
= 1
+ \alpha^{\,\prime}(m) \, \rho \, \cos (\phi -\delta)
+ \beta^{\prime}(m) \, \rho^{2} \, \cos 2\left(\phi -\delta\right),
\label{EQU:Dl51}
\eeq
with
\beq
\delta=\arctan\frac{2\eta\sqrt{1-\mu}}{\mu}, \qquad
\rho =\sqrt{1+4\eta^2\,\frac{1-\mu}{\mu^2}}, \qquad
\mu = \left(2m_V/m\right)^2 < 1.
\label{EQU:delro}
\eeq
Measurement of this rotation $\delta$ of the azimuthal distributions,
would demonstrate $CP$ violation in the coupling between the Higgs
boson and the vector bosons.
In order to facilitate such a possibility, we introduce the following
measures of the asymmetry:
\beqa
\left.
\begin{array}{c}
A\left(m\right) \\
A^{\prime}\left(m\right)
\end{array}
\right\}
& \equiv &
\frac{1}{\pi} \int_{0}^{2\pi}\dd\phi
\left\{
\begin{array}{c}
\cos \phi \\
\sin \phi
\end{array}
\right.
\left(\frac{2 \pi}{\Gamma}\:\frac{\dd\Gamma}{\dd\phi}\right) =
\rho \ \alpha^{\,\prime}\left(m\right)
\left\{
\begin{array}{c}
\cos \delta \\
\sin \delta
\end{array}
\right.,
\label{EQU:nr2} \\
\left.
\begin{array}{c}
B\left(m\right) \\
B^{\prime}\left(m\right)
\end{array}
\right\}
& \equiv &
\frac{1}{\pi} \int_{0}^{2\pi}\dd\phi
\left\{
\begin{array}{c}
\cos 2\phi \\
\sin 2\phi
\end{array}
\right.
\left(\frac{2 \pi}{\Gamma}\:\frac{\dd\Gamma}{\dd\phi}\right) =
\rho^{2} \ \beta^{\prime}\left(m\right)
\left\{
\begin{array}{c}
\cos 2\delta \\
\sin 2\delta
\end{array}
\right.,
\label{EQU:nr1}
\eeqa
where the unprimed observables
correspond to the $SM$ prediction (modulo corrections of
order $\eta^{2}$)
and the primed ones correspond to the $CP$ violating contributions.
This identification is evident when we note that, for any $\eta$,
\beq
\frac{2 \pi}{\Gamma}\:\frac{\dd\Gamma}{\dd\phi} =
1 + A\left(m\right) \cos \phi +B\left(m\right) \cos 2\phi
  + A^{\prime}\left(m\right) \sin \phi+B^{\prime}\left(m\right) \sin 2\phi.
\label{EQU:nr5}
\eeq
Presence of $CP$ violation will now manifest itself as a non-zero value
for the observables $A^{\prime}$ and $B^{\prime}$; the magnitude of
$CP$ violation may be determined from the angle $\delta$,
\beq
\tan \delta = \frac{A^{\prime}\left(m\right)}{A\left(m\right)} \qquad
\mbox{\rm and} \qquad
\tan 2\delta  = \frac{B^{\prime}\left(m\right)}{B\left(m\right)}.
\label{EQU:nr6}
\eeq
In principle, this allows for consistency checks, since the two expressions in
eq.~(\ref{EQU:nr6}) are closely related. However, a possible experimental
observation of $CP$ violation is strongly dependent upon the magnitudes
of the observables $B^{\prime}$ and $A^{\prime}$.
A first question in this direction is: which of them is easier to detect?
In order to answer this question, we may study the ratio
\beq
\frac{A^{\prime}\left(m\right)}{B^{\prime}\left(m\right)} =
\frac{\alpha^{\,\prime}\left(m\right)}{2 \beta^{\prime}\left(m\right)}.
\label{EQU:forh1}
\eeq
This ratio is {\it independent of $\eta$}.
In the case of $H \rightarrow W^{+}W^{-}\rightarrow 4 f$,
$A^{\prime}$ is relatively important for any Higgs mass. In the case of
$H \rightarrow ZZ\rightarrow 4 l$, $B^{\prime}$ is relatively
important for intermediate Higgs masses, whereas $A^{\prime}$ becomes
more important for $m\gsim$~600~GeV.
(The case $H \rightarrow ZZ \rightarrow 2l\,2q$ is intermediate.)

The amplitude functions $\alpha^{\,\prime}(m)\rho$ and
$\beta^{\prime}(m)\rho^2$
of (\ref{EQU:Dl51}) are given in
\cite{osskj2} for the cases
$H \rightarrow W^{+}W^{-}\rightarrow 4 f$ and
$H \rightarrow ZZ\rightarrow 4 l$, respectively.
We note that $\beta^{\prime}(m)\rho^2$ is independent of the relative
strengths of axial and vector couplings, whereas $\alpha^{\,\prime}(m)\rho$
is proportional to the $\sin2\chi_i$ factors.
Both amplitudes are small in most of the
$\left(\eta, m, \sin 2\chi_{i}\right)$ parameter space,
but $\alpha^{\,\prime}(m)\rho$ is comparable to unity
for any value of $\eta$,
in the intermediate Higgs mass range in the case
$H \rightarrow W^{+}W^{-}\rightarrow 4 f$.
The angle $\delta$ is given in Fig.~\ref{plfun1} (left)
for different values of $\eta$.
Provided $\eta$ is not too small,
this angle is significantly different
from zero when the Higgs is well above threshold.
It appears that an experimental observation
of $CP$ violation would only be possible in restricted ranges of
$\eta$ and Higgs mass values, and only for selected decay channels.
\begin{figure}[t]
\begin{center}
\setlength{\unitlength}{1cm}
\begin{picture}(16,7.0)
\put(-0.5,-2.5){\mbox{\epsfysize=11cm\epsffile{cpvio1.eps}}
             \hspace*{-10mm}
             \mbox{\epsfysize=11cm\epsffile{cpcb.eps}}}
\end{picture}
\vspace*{15mm}
\end{center}
\caption{\label{plfun1}{\it Left} : The angle $\delta$ (in degrees) for a Higgs
particle of mass $m$, for $\eta=1$, $10^{-1}$, $10^{-2}$, and $10^{-3}$.
{\it Right} : The ratio $C^{\,\prime}\left(m\right)/B^{\prime}\left(m\right)$
for
a Higgs of mass $m$ in the cases of $H \rightarrow W^{+}W^{-} \rightarrow 4 f$
and $H \rightarrow ZZ \rightarrow 4 l$, for $\eta=10^{-3}$, $10^{-1}$, and
$1$.}
\end{figure}

Let us now turn to a discussion of the decay rate weighted
with energy differences.
In the narrow-width approximation we obtain
\beq
\frac{2 \pi}{\tilde{\Gamma}}
\:\frac{\dd\tilde{\Gamma}}{\dd\phi}
 =
1 + \frac{\kappa^{\prime}\left(m\right)}{\rho } \cos (\phi-\delta),
\label{EQU:Dl71}
\eeq
with
\beq
\kappa^{\prime}(m) =
\frac{1}{\left(4 \sin 2 \chi_{1} \sin 2 \chi_{2}\right)^{2}}
\frac{\alpha^{\,\prime}(m)}{\beta^{\prime}(m)},
\label{EQU:Dkappa}
\eeq
which is independent of $\eta$.
As in eqs.~(\ref{EQU:nr1})--(\ref{EQU:nr2}) we introduce the
following measures of asymmetry:
\beq
\left.
\begin{array}{c}
C\left(m\right) \\
C^{\,\prime}\left(m\right)
\end{array}
\right\}
\equiv
\frac{1}{\pi} \int_{0}^{2\pi}\dd\phi
\left\{
\begin{array}{c}
\cos \phi \\
\sin \phi
\end{array}
\right.
\left(\frac{2 \pi}{\tilde{\Gamma}}\:\frac{\dd\tilde{\Gamma}}{\dd\phi}\right) =
\left(\frac{\kappa^{\prime}\left(m\right)}{\rho}\right)
\left\{
\begin{array}{c}
\cos \delta \\
\sin \delta
\end{array}
\right.,
\label{EQU:nr02}
\eeq
so that
\beq
\frac{2 \pi}{\tilde{\Gamma}}
\:\frac{\dd\tilde{\Gamma}}{\dd\phi}
 =
1 + C\left(m\right) \cos \phi
+ C^{\,\prime}\left(m\right) \sin \phi.
\label{EQU:Dlk7}
\eeq
Consequently
\beq
\tan \delta  = \frac{C^{\,\prime}\left(m\right)}{C\left(m\right)}.
\nn
\eeq
Hence, this set of observables provides yet another way of measuring
the $CP$ violating phase $\delta$.
Although the ratio of these energy-weighted
observables coincides with, or is trivially related to the previous ones,
the possibility of demonstrating a presence of $CP$ violation has increased
significantly. The ratio
\beq
\frac{C^{\,\prime}\left(m\right)}{A^{\prime}\left(m\right)} = \frac{1}{\left(4
\sin 2 \chi_{1} \sin 2 \chi_{2} \right)^{2} \beta^{\prime}\left(m\right)
\rho^2}
\label{EQU:forh2}
\eeq
is given in Fig.~\ref{plfun1} (right) for $\eta=10^{-3},
10^{-1}$, and $1$ in the
cases of $H \rightarrow W^{+}W^{-}\rightarrow 4 f$ and
$H \rightarrow ZZ\rightarrow 4 l$.
We see that the energy-weighted
$CP$-violating observable $C^{\,\prime}$ is
a much more sensitive probe for
establishing $CP$ violation than the former ones. The amplitude
function $\kappa^{\prime}(m)/\rho$ of (\ref{EQU:Dl71}) is given in
\cite{osskj2} for the cases
$H \rightarrow W^{+}W^{-}\rightarrow 4 f$ and
$H \rightarrow ZZ\rightarrow 4 l$.
This amplitude turns out to be comparable to,
or much bigger than unity for arbitrary values of $\eta$
and for Higgs decay to any observable four-fermion final state.

\section{Summary and conclusions}
\setcounter{equation}{0}
\label{sec:conc}

We notice that the special cases $\eta=0$ and $\eta \gg 1$ in the
former section correspond to the $CP$ even and $CP$ odd eigenstates,
respectively. Hence, the distributions (\ref{EQU:Dl51}) and (\ref{EQU:Dl71})
should be interpreted as being intermediate to the two eigenstates; see
Fig.\ref{plfun2}.

We see that the energy-weighted distribution is generally
a much more sensitive probe for $CP$ than the
purely angular distribution of eq.~(\ref{EQU:intro1}).
In addition, for $V=Z\rightarrow 2l$ the $\sin 2\chi$-factors
provide an enhancement in the energy-weighted amplitude function, and
it is encouraging that this enhancement occurs for the so-called
``gold--plated mode'' $H \rightarrow ZZ\rightarrow
e^{+}e^{-}\mu^{+}\mu^{-}$ \cite{HHG}.


\medskip
It is a pleasure to thank P. Osland for useful discussions.
I am grateful to the Organizers of the Zvenigorod Workshop,
in particular Professor V. Savrin,
for creating a very pleasant atmosphere during the meeting.
This research has been supported by the Research Council of Norway.

\vspace{8mm}

\end{document}